\documentclass[a4paper]{jpconf}
\usepackage{graphicx}
\begin{document}
\title{Atomic Processes in Planetary Nebulae and H~{\sc ii} Regions}

\author{Manuel A. Bautista}

\address{Centro de F\'{\i}sica, IVIC, P.O. Box 21827 Caracas 1010A, Venezuela\footnote{Currently at Department of Physics, Virginia Polytechnic Institute and State University, VA 24061, USA}}

\ead{bautista@vt.edu}

\begin{abstract}
Spectroscopic studies of Planetary Nebulae (PNe) and  H~{\sc ii} regions have driven much    
development in atomic physics. In the last few  years the combination of a 
generation of powerful observatories, the development of ever more 
sophisticated spectral modeling codes, and large efforts on mass 
production of high quality atomic data have led to important progress in our 
understanding of the atomic spectra of such astronomical objects. In this paper I review such 
progress, including evaluations of atomic data by comparisons with nebular spectra,
detection of spectral lines from most iron-peak elements and n-capture elements, 
observations of hyperfine emission lines and analysis of isotopic 
abundances, fluorescent processes, and new techniques 
for diagnosing physical conditions based on recombination spectra. 
The review is directed toward atomic physicists and spectroscopists 
trying to establish the current status of the atomic data and models 
and to know the main standing issues.
\end{abstract}

\section{Introduction}
Nearly all modern observational research in gaseous nebulae (PNe, H~{\sc ii}
regions, and circumstellar nebulae) involves some kind
of spectra, whose interpretation requires some understanding of the atomic
processes involved and the handling of atomic data. The level of understanding                     
of such processes and the data required for various applications ranges from                          
phenomenological and statistical studies of commonly prominent lines, 
to compilation of raw atomic data for direct inspection
(such as line finding lists), and to modeling of synthetic spectra attempting                            
to fit the observed spectra. On the other hand, the accuracy and quantity
of atomic models describing various features in observed spectra have
evolved with time, driven by great advances in ground- and space-based
observatories, computer technology and experimental techniques. At present
there are models and data accurate enough for satisfactory analysis of at
least the most prominent spectral features. Also the advent of
on-line databases and spectroscopic tools has revolutionized the                                              
dissemination of results of atomic physics research. However, 
the demands on the quantity and quality of the atomic data will grow as                                  
new instruments delivering greater sensitivity and spectral resolution
become available. 

A review of atomic processes spans an audience that includes specialists
in the study of atomic systems with tools capable of providing models and
atomic data, nebular astrophysicists, astronomers seeking to understand the reliability and
accuracy of their modeled spectra, and those who use the models and data
to compare synthetic spectra with observations, diagnose the physical
conditions and compute chemical abundances of nebulae. The present
review is aimed at researchers in atomic physics and spectroscopy who wish
to understand the reach of currently available atomic models and identify 
present needs.

The physical conditions in gaseous nebulae of interest here are those
of photoionized plasmas as observed in the ultraviolet (UV), 
optical, and infrared (IR)
bands. Roughly speaking these conditions are: temperatures of the order of 
$10^4$~K and electron densities between $10^3$ and $10^8$~cm$^{-3}$.
This review will not treat the X-ray band which in gaseous nebulae
results from mechanically heated coronal gas.                                                                             
A good review on this subject can be found in
\cite{kallman}. 

The present review is organized according the nuclear charge of atomic
species, starting with hydrogen and helium, and following with the second
and third row elements, the iron-peak ions, and then onto the heavier
elements. I then finish the paper with a discussion of various general
issues of interest.

\section{Hydrogen}

The spectrum of hydrogen in H~{\sc ii} regions has been studied extensively for quite some time.
In principle the spectrum is determined by the recombination rates into each 
level and the subsequent radiative cascades. Such a process is accurately 
modeled in two idealistic situations, the so called ``case A'' in which 
 all lines are optically thin and ``case B" which assumes that the
optical depth of Ly$\alpha$ goes to infinity. Two complications arise
beyond these approximations: (1) a detailed solution to the radiative 
transfer
of Ly$\alpha$ photons including the effects of self-absorption, removal
and collisionally induced transitions between the $2s$ and $2p$ states,
and (2) when the electron temperature and density of the plasma are high
enough for effective collisional excitations on to $n>2$ states.                                                  

Dennison et al. \cite{den05} 
discuss the first of these problems and 
propose an observational test. They claim that radio-observatories will  
soon be able to detect the $2s_{1/2}-2p_{1/2}$ and $2s_{1/2}-2p_{3/2}$ lines at
1.1 GHz and 9.9 GHz respectively. Dennison et al. explain 
that removal of Ly$\alpha$ photons by dust would limit the pumping of H atoms 
into the $2p$ states, and under these conditions the $2s_{1/2}-2p_{1/2}$ 
transition will appear in stimulated emission while the $2s_{1/2}-2p_{3/2}$                            
line will appear in absorption. In general, the relative strengths of these 
two lines should serve as diagnostics of the populations of the 
$2s$ and $2p$ states and the responsible processes. Further, 
Dennison et al. suggest that removal by dust is the dominant mechanism acting 
on Ly$\alpha$ photons in H~{\sc ii} regions, which if correct would circumvent the need to solve the radiative transfer problem.

In regards to collisional excitation of hydrogen, 
P\'equignot \& Tsamis \cite{peq05}  
study the evolution of calculated collision strengths for $1s \to
n=3, 4, 5$. They point out large variations in the theoretical values up
to the last calculation of Anderson et al. \cite{and00}. P\'equignot \& Tsamis
compare their observations for the PN G135.9+55.9 with predictions of
models using various collisional data sets. They find that the collision
strengths of Anderson et al. yield the most consistent results, yet these
may have not reached ultimate accuracy.

Modern high-signal-to-noise optical spectra allow for accurate measurements of                       
lines and continuum around the Balmer and Paschen recombination series.                                
By fitting the results of detailed spectral models to these measurements 
it is possible to determine the electron density and temperature in the nebula                             
\cite{zha04}. Surprisingly, temperatures obtained from the Balmer series
often disagree with the temperatures derived from line ratios of collisionally excited lines, such as the ratio of nebular to auroral lines of [O~{\sc iii}].
Moreover, for a large sample of PNe studied by Zhang et al., the hydrogen                             
Balmer temperatures are typically lower than the collisional oxygen temperature
by several thousand degrees. These temperature differences are used in support 
of the so called 'temperature fluctuations' or 'temperature variations' in
nebulae. However, a flag of warning on the use of hydrogen temperatures comes
from the fact that in four PNe where temperatures from the Paschen series were     
determined, these disagree with the Balmer temperatures at the three sigma                              
level or more.

\section{Helium}
  
Modeling the He~{\sc i} recombination spectrum has become increasingly
reliable, with an accuracy on line emissivities of the order of 
a few percent
(see \cite{bau05,ben99,ben02}). Yet, this is not at the 1\% accuracy level needed to place 
useful constrains on the primordial helium abundance in cosmological studies \cite{oli04,izo04}.                                                                        
Recently, Bauman et al. \cite{bau05} worked out the recombination problem in fine structure
in the low density limit. They found no significant effects on the 
recombination spectrum from spin-orbit coupling. However, they found
that some of lines may be uncertain due to use of poor quality atomic                                        
data, specifically the photoionization cross sections for states with
$9<n<20$ and $L<3$. 

A practical test on the accuracy of models and observations of He~{\sc i}
lines came from Porter et al. \cite{por07} who compared model predictions with measured
fluxes for 100 lines with $n\le 20$ from the Orion nebula. They found
an average difference of 6.5\% between all observed and predicted lines 
and 3.8\% difference for the 22 most accurately measured lines. 

Zhang et al. \cite{zha05a,zha05b} present a method to diagnose temperatures from He~{\sc i}
line ratios. They write 
parametric forms for He~{\sc i} line ratios vs. temperature and conclude
that the best diagnostic is obtained from the $\lambda7281/\lambda6678$
ratio. The results of the diagnostics have an intrinsic uncertainty of
$\sim$1000~K at around 10~000~K owing to optical depth effects on the
He~{\sc i} $\lambda3889$ ($2s~^3S \to 3p~^3P^o$) line. When comparing the
results from the $\lambda7281/\lambda6678$ ratio with those from
$\lambda7281/\lambda5876$ a systematic error of ($\sim$500~K) arises,
which may be related to the uncertainty on the optical depth of the
$\lambda3889$ line, or perhaps from some other source. Interestingly, in a
sample of 48 PNe the helium temperature is systematically lower than that
obtained from the hydrogen H$\beta$ decrement by an average of 4000~K.

\section{Second and third row elements (C-Ar)}

Density and temperature diagnostics in PNe are commonly based on line
ratios among collisionally excited lines. The best known density
diagnostics are [S~{\sc ii}] $\lambda6717$/$\lambda6731$ and [O~{\sc ii}]
$\lambda3729/\lambda3726$ that are sensitive to densities of the order of
$10^3$~cm$^{-3}$, [Cl~{\sc iii}] $\lambda5517/\lambda5537$ for $N_e\sim
10^4$~cm$^{-3}$, and [Ar~{\sc iv}] $\lambda4711/\lambda4740$ for $N_e\sim
10^4-10^5$~cm$^{-3}$.

The [O~{\sc ii}] ratio that arises from transitions among the $^4S_{3/2}$
ground level and the $^2D_{5/2}$ and $^2D_{3/2}$ levels was a subject of
controversy for the last few years. It was commonly assumed that
$LS$-coupling was a valid approximation for terms of the ground
configuration of the O$^+$ system; thus the ratio of collision strengths                                
from the ground levels to the $^2D_{5/2}$ and $^2D_{3/2}$ levels was given
by the statistical weights, i.e. 1.5. However, McLaughlin \& Bell \cite{mcl98}
claimed that relativistic effects enhanced the ratio of 
collision strengths to 1.93, with profound effects on $N_e$ determinations
for low surface brightness H~{\sc ii} regions. Such claims were contested
by evidence from the extensive literature survey of Copetti \& Writzl \cite{cop02} 
and the observational campaign of Wang et al. \cite{wan04}. The issue
seems settled now with a new calculation by Pradhan et al. \cite{pra06}, that
accounts for all dominant relativistic effects and confirms the earlier
predictions of the $LS$-coupling approximation. The spectroscopic survey by
Wang et al.  also served to test the $A$-values for dipole                                                   
forbidden transitions of [O~{\sc ii}]. They found that the transition                                        
probabilities of Zeippen \cite{zei82} best fit the observations, the
differences being of only a few percent, while the results of later
calculations by the same author (\cite{zei87}) and by W. Wenaker \cite{wwen90} look  
problematic. This latter set is what is currently available through the NIST database
(\cite{wie96}).

Wang et al. also compared the electron densities obtained from various
line ratios and obtained very good agreement between $N_e$([O~{\sc ii}]),
$N_e$([S~{\sc ii}]), and $N_e$([Cl~{\sc iii}]). By contrast $N_e$([Ar~{\sc
iv}]) values yield densities systematically higher than those from the
other diagnostics, which sheds some doubts on the accuracy of the
$N_e$([Ar~{\sc iv}]) $A$-values.                                                                                     

\section{The iron-peak elements}

Fe and iron-peak elements are important constituents of gaseous nebulae. Depending on
the excitation of the nebula, iron is frequently seen in stages from
Fe~{\sc i} to Fe~{\sc vii}. The observed ions of nickel span about the
same range. Other less abundant elements of the group are occasionally
identified as well. An extreme example of rich spectra is the $\eta$~Carinae
ejecta (see article by Gull in this volume).

On this subject the contributions of Sveneric Johansson could hardly be
overstated. He has been the driver of extensive and detailed research                                   
of energy levels and transition rates for the low ionization stages
of iron-peak species. Data often are taken for granted as they becomes                           
easily available through various databases, yet their determination                                       
from laboratory work requires very skillful people and substantial
funding. Johansson's publications on the subject are too numerous to list
here, but some of the most important contributions are:
on the spectra of Sc~{\sc ii} \cite{joh80}, 
Fe~{\sc i} \cite{nave94}, Fe~{\sc ii} \cite{joh78},
Ti~{\sc ii} \cite{hul82}, Co~{\sc ii} \cite{pic98}, and on the 
determination of $f$-values, lifetimes, and radiative rates for                                          
forbidden transitions (e.g. for Fe~{\sc ii} \cite{ber96,sik99,ros01,joh77a},
Ti~{\sc ii} \cite{har03}, Ni~{\sc i} \cite{joh03}). 
Johansson has also done seminal work the understanding of
the Ly$\alpha$ excitation mechanism of Fe~{\sc ii} and other
iron-peak species in astronomical spectra (e.g. \cite{joh84,joh77b,har00,
joh93}).

Spectra of  singly ionized iron-peak species are particularly complex
as they respond to a variety of excitation mechanisms.
These are H Ly$\alpha$ fluorescence, continuum fluorescence, 
self-fluorescence by 
 overlapping Fe~{\sc ii} transitions, and collisional 
transitions among low and high levels. 
In moderately dense plasmas ($N_e \sim10^7$~cm$^{-3}$) the populations of high levels involved in fluorescence are 
redistributed by collisionally induced transitions through highly 
excited pseudo-metastable levels \cite{bau04b}.
Furthermore, proper modeling of Fe~{\sc ii} spectra requires large systems, approaching one thousand levels and complete and accurate atomic data.
This has led to a diverse set of models 
with little consensus among them and limited understanding about their accuracy.
For ions of atoms other than Fe and Ni the first detailed spectral models are
now being created, e.g. \cite{bau06,bau07}. 
A good part of this work has been done under the auspices of the                                              
IRON Project \cite{ironp}. 
A summary of the data is presented in Fig.~1.

More work is also in progress  
by various groups like the IRON Project, the FERRUM
Project at Lund University, at NIST, and the group at Queen's University                     
Belfast.    

\begin{figure}
 \includegraphics[height=4in,width=6in,angle=+00]{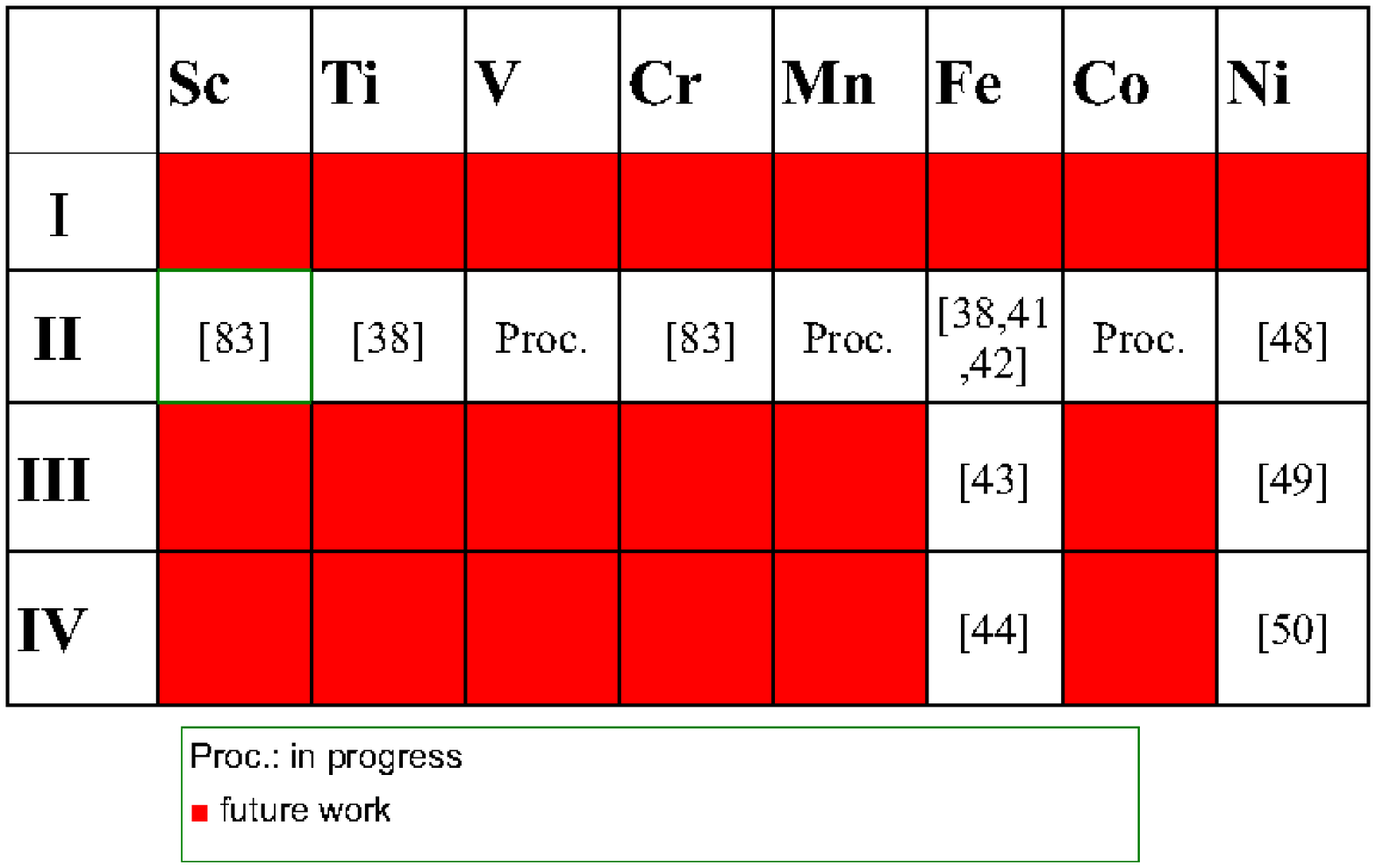}
  \caption{Spectral models for low ionization
iron-peak species}\label{fig1} 
\end{figure}

Another modeling challenge in dealing with dense spectra of iron-peak species
is the treatment of the transfer of line radiation from the point of emission to                          
the boundary of the atmosphere. For iron-peak species, in addition to 
continuum opacity and resonant scattering, one must consider fluorescence by 
line overlaps, such as the well known H Ly$\alpha$ñ effect in Fe~{\sc ii} \cite{joh84}.                                                                                                 
While writing the present review, I have estimated the number of overlaps between absorption lines arising from 
low excitation levels, typically populated under nebular conditions, of 
Fe~{\sc ii-iv} and lines from H~{\sc i}, He~{\sc ii}, O~{\sc ii} and O~{\sc iii}, dominant species in typical nebular spectra. Line wavelengths awere taken from 
the NIST database and do not represent the entire spectra. 
For line widths of 20~km/s the number of overlaps
are 592, 243, and 49 
for Fe~{\sc ii}, {\sc iii}, and {\sc iv} respectively,
while for widths or 60~km/s the overlaps are 1722, 762, 524 for the same
ions.  
This is the range of line widths in nebular spectra.
The huge number of line overlaps 
demonstrates the need for a detailed treatment of radiative transfer of Fe 
spectra. The case also applies to other iron-peak species. 

In modeling spectra one also needs to solve for ionization equilibrium of each
element. For that reason, detailed level specific photoionization cross sections, that 
account for complex resonance
structures, are being
computed for both ground and excited states using the R-matrix method. Cross sections are available for Fe~{\sc i}-{\sc iv}
\cite{bau97,nah94,nah96,bau97b} and Ni~{\sc ii} \cite{bau99}.
Recently, much progress has been made at the {\it Instituto Venezolano de Investigaciones Cientificas} (IVIC) on the calculation of cross  sections
for neutral Sc, Ti, and Cr. 

\section{Hyperfine-induced spectra}

Observations of hyperfine-induced transitions in PNe have been of great interest in recent years. This is because these spectra allow us to determine relative
isotopic compositions to be compared with predictions of 
nucleosynthesis. This, in turn, enables us to disentangle enrichment by stellar nucleosynthesis
from the primordial composition in observed present total abundances.                                     
For example, the abundance of $^3$He in our Galaxy could be used to test 
big bang nucleosynthesis and constrain the baryonic density of the Universe.
However, the evolution of $^3$He in the Galaxy is a longstanding open problem
due to the observed discrepancy (by two orders of magnitude) 
between theoretical yields of low-mass stars and the measured abundances in 
H~{\sc ii} regions \cite{roo76,tos01}. A proposed solution of this problem                               
could be in suppressing, by non-standard mixing mechanisms,  the production of $^3$He during the red giant branch 
and/or asymptotic giant branch phases of stars of mass up to $\sim$2 $M_\odot$
\cite{hog95,cha95}. 
An observable consequence  of such an scenario is that the ratio $^{12}$C/$^{13}$C              
in the ejecta of PNe should be much lower than in the standard case.
For a 1 $M_\odot$ star, the predicted ratio is $\sim$5, in contrast with the 
standard ratio of 25-30 \cite{boo99}.

Isotopic abundances can be derived from beryllium-like ions through observations
of the hyperfine-induced transition 2s2p~$^3P^0_0~-~$2s$^2~^1S_0$ within the 
UV0.01 multiplet. This transition can only occur through coupling of the electronic total momentum with nuclei of nonzero spin. This
happens in the spectra of $^{13}$C (with nuclear spin 1/2) but not in
$^{12}$C. Thus, $^{12}$C/$^{13}$C abundance ratios can be derived from 
spectra showing the hyperfine transitions together with the stronger
2s2p~$^3P^0_2~-~$2s$^2~^1S_0$ (M2) and 2s2p~$^3P^0_1~-~$2s$^2~^1S_0$ (IC)
transitions. Palla et al. \cite{pal02} observed these lines in NGC~3242 
using STIS on board the {\it Hubble Space Telescope}. From this, they derived
$^{12}$C/$^{13}$C$\>$=38. The same lines from the isoelectronic ion N~{\sc iv}                
were measured by Brage et al. \cite{bra02} from a STIS spectrum of                                                                                                              
NGC~3918. In this case both stable isotopes of nitrogen ($^{14}$N and 
$^{15}$N) have non-zero nuclear spin. Although no abundances were derived, this work served to confirm
the theoretical $A$-value for the hyperfine transitions.                                                           
More recently, Rubin et al. \cite{rub04} analyzed spectra of
PNe in the  IUE archive and determined $^{12}$C/$^{13}$C $= 4.4\pm1.2$ for one               
object and established lower limits to this isotope ratio in 23 objects.

Another study of hyperfine induced transitions, but this time in Ne-like
Al~{\sc iv}, was reported by Casassus et al. \cite{cas05}. The two stable 
isotopes, $^{27}$Al and $^{26}$Al, have nuclear spins 5/2 and 5 respectively. Thus,
hyperfine structure causes the $^3$P$_2$~-~$^3$P$_1$ 
transition at 3.66$\mu$m 
to split into 9 components, 5 of which could be resolved 
spectroscopically. From the observations, Casassus et al. calculated an upper
limit to the $^{26}$Al/$^{27}$Al abundance ratio of 1/33 in NGC~6302, 
which is in
good agreement with the predicted value of 1/37 from stellar evolution models.

\section{Neutron capture elements}

One of the frontiers of spectroscopy of PNe is the study of {\it
n}-capture elements presumably synthesized by the AGB progenitor of the
nebula. The progress on this area since the seminal work of 
P\'equignot \& Baluteau \cite{peq94}
has been slow owing to the need for high spectral
resolution ($>$ 10~000) and the absence of atomic data and models. Despite
the difficulties, Dinerstein  \cite{din01}                                                                                    
was able to identify lines of 
[Kr~{\sc iii}] and [Se~{\sc iv}] in the K-band of the near infrared spectra of
NGC~7027 and IC~5117. Soon after, Dinerstein \& Geballe \cite{ding01} also
detected [Zn IV] lines in the near infrared spectrum of
NGC~7027 . To date these lines have been identified in almost one
hundred PNe \cite{ste05,ste07}.

On the theoretical side, the calculation of collisional data for heavy
species represents an important challenge because intermediate coupling
representations are inappropriate in many cases, at the same time as
relativistic effects become too large to treat in the Breit-Pauli
representation. Thus, researchers must turn to fully relativistic codes
in the Dirac formalism. Recently, Badnell et al. \cite{bad04} put
the Dirac Coulomb R-matrix package of Berrington et al. \cite{ait96} on the same
footing as the traditional Breit-Pauli R-matrix codes developed by the
IRON/RmaX team. Nonetheless, these sort of calculations are still far from
routine work, owing to the size of the atomic representations in $JK$
-coupling.

\section{Additional considerations}

This review would not be complete without pointing out some important
general issues of relevance to the present subject as well as recently
developed tools for atomic spectroscopy.

The accuracy of recombination rate coefficients was in recent years an
important source of concern for modelers for their effects on calculations
of ionic fractions. The issue, however, seems to have reached a point of
stability as different theoretical methods, e.g. the unified R-matrix
approach of Nahar and Pradhan and the Thomas-Fermi-Dirac approach in the
version of Badnell and collaborators \cite{bad03}, begin to yield
consistent results and to release large amounts of data. Such theoretical
results also seem to agree with recent experimental determinations (e.g.                  
\cite{fog05}). A warning must be raised though regarding low
temperature dielectronic recombination, that is dominated by near
threshold autoionizing channels, whose positions are difficult to get
accurately from theory. More experimental work on this area is needed to
benchmark current and forthcoming data (e.g. \cite{bohm05}).

On its origins, one of the explicit objectives of the IRON Project was to
provide a complete and reliable set of data for positive ions of interest
in nebular infrared spectroscopy. A comprehensive review of these data is
presented by Badnell et al. \cite{bad06}. The entire
dataset produced by the IRON Project will soon be available through
TIPbase \cite{men00}.

A very useful tool for nebular spectroscopy, the EMILI package for emission 
line identification, has been developed by \cite{sha03}. This tool should 
expedite the analysis of high spectral resolution and signal-to-noise spectra, and enable the identification of faint poorly known features. 

\section{Conclusions} 

The question of what fractions of Ly$\alpha$ photons are self-absorbed and
removed by dust in photoionized regions is rather important, as the
treatment of hydrogen spectra in photoionization models is one of the                                     
main limitations to their accuracy. There are also various questions in
regard to heating and photo-evaporation of grains in PNe. Further
research on this 
is fundamental for entering a new level of
detailed understanding of PNe and H~{\sc ii} regions. 

Temperature diagnostics with H and He are very important as they yield
new information on the long lasting problem of temperature fluctuations
in gaseous nebulae. It is important, though, to sort out the apparent
inconsistencies between temperature diagnostics from different recombination
series of lines or line ratios. Moreover, it seems plausible that temperature
fluctuations in nebulae could be accompanied by density variations. Thus, it                          
is important that both quantities be derived simultaneously from the
same diagnostics. 

As atomic models have become available for essentially all of the most
prominent species in spectra of gaseous nebulae it is important to check that 
diagnostics of temperature and density out of various species all yield 
consistent results. This would allow us to determine whether the atomic data
have reached ultimate accuracy. A flag of warning must be raised, however,
in regard to neutral species and ions with ionization potentials below
that of hydrogen because their spectra could be affected by
photo-excitation by continuum radiation. Such is the case of N~{\sc i}
\cite{bau99b} that Copetti \& Writzl \cite{cop02} tried to analyze on the same
grounds as ionized species.

Iron and iron-peak ions are as important as ever, particularly Fe~{\sc
ii}, which is prominent throughout astronomy, beyond PNe research. These
ions, however still pose great challenges.
The open 3d structure of these species is difficult to
describe owing to various factors: (1) large numbers of strongly correlated
$LS$ terms shortly spaced in energy from the ground state; (2) strong
angular correlations among configurations such as 4s$^2$, 4p$^2$, 4d$^2$,
4p4f that lead to very demanding computations; (3) strong core-valence
interactions in the vicinity of excitations of 3p electrons onto the
unoccupied 3d orbital, leading to the so-called giant
3p$\to$3d resonance; 
(4) electronic correlations among 3d electrons such as
the radial distribution of electrons in a 3d$^{N+1}$ configuration
differs from that of electron in a 3d$^N$nl. 
More experimental work is needed, particularly in the measurement of
cross sections that guide theoretical work (e.g. \cite{bau06b}).

The importance to modern astronomy of studying isotopic abundances and
$n$-capture elements can hardly be overstressed. Yet, such work imposes
extreme demands on spectroscopic instruments and astronomers. Carefully
coordinated efforts between atomic physicists, astronomers, and instrument
designers would be desirable on this subject.

\verb"\ack"
\subsection{Acknowledgments}
I wish to thanks the referee of this paper, Prof. D. Morton, 
important discussions and corrections to the original manuscript.


\section*{References}

\begin{thebibliography}{9}

\bibitem{kallman} Kallman, T.R. \& Palmeri, P. 2006 {Ann. Rev. Mod. Phys.} 79 79

\bibitem{den05} Dennison, B., Turner, B.E. \& Minter, A.H. 2005 {Astrophys. J.} 633 309
\bibitem{peq05} P\'equignot, D. \& Tsamis, Y.G. 2005 {Astron. \& Astrophys.} 430 187
\bibitem{and00} Anderson, E., Ballance, C.P.,Badnell, N.R., \& Summers, H.P. 2000 {J. Phys. B: Opt. Atom. Mol. Phys.} 33  1255

\bibitem{zha04} Zhang, Y., Liu, X.-W., Wesson, R., Storey, P.J., Liu,
Danzinger, I.J.  2004  {Mon. Not. Royal Astron. Soc.}  351  935

\bibitem{bau05} Bauman, R.P., Porter, R.L., Ferland, G.J.\& MacAdam, K.B. 2005  {Astrophys. J.} 628  541
\bibitem{ben99} Benjamin, R.A., Skillman, E.D. \& Smits, D.P. 1999  {Astrophys. J.} 514  307
\bibitem{ben02} Benjamin, R.A., Skillman, E.D. \& Smits, D.P. 2002  {Astrophys. J.} 569  288
\bibitem{oli04} Olive, K.A. \& Skillman, E.D. 2001  {New Astron.} 6  119
\bibitem{izo04} Izotov, Y.I. \& Thuan, T.X., 2004  {Astrophys. J.} 602  200
\bibitem{por07} Porter, R.L., Ferland, G.J., MacAdam, K.B. 2007  {Astrophys. J.} 657  327
\bibitem{zha05a}{Zhang, Y., Liu, X.-W., Liu, Y. \&
Rubin, R.H.} 2005a  { Mon. Not. Royal Astron. Soc.} 358  457

\bibitem{zha05b} Zhang, Y., Rubin, R.H. \&
Liu, X.-W. 2005b  {Rev. Mex. AA (Serie de Conferencias)} 23  15
\bibitem{mcl98} McLaughlin, B.M. \& Bell, K.L. 1998  {J. Phys. B: Atom. Mol. Phys.} 31  4317

\bibitem{cop02} Copetti, M.V.F. \& Writzl, B.C. 2002  {Astron. \& Astrophys.} 382  282

\bibitem{wan04} Wang, W., Liu, X.-W., Zhang, Y. \& Barlow, M.J. 2004 
{ Astron. \& Astrophys.} 427  873

\bibitem{pra06} Pradhan, A.K., Montenegro, M., Nahar, S. \& Eissner, W. 2006  {Mon. Not. Royal Astron. Soc.} 366  L6

\bibitem{zei82} Zeippen, C.J. 1982  { Mon. Not. Royal Astron. Soc.} 198  111

\bibitem{zei87} Zeippen, C.J. 1987  { Astron. \& Astrophys.} 173  410

\bibitem{wwen90} W. Wenaker, I., 1990  {Phys. Scr.} 42  667

\bibitem{wie96} Wiese, W.L., Fuhr, J.R. \& Deters, T.M. 1996, {Atomic transition probabilities for carbon, nitrogen, and oxygen: a critical data compilation} Washington, Am. Chem. Soc. for the N.I.S.T.
\bibitem{joh80} Johansson, s. \& Litz\'en, U. 1980  {Phys. Scripta} 22  49

\bibitem{nave94} Nave, G., Johansson, S., Learner, R. C. M., Thorne, A. P., Brault, J. W. 1994, {Astrophys. J. Sup.} 94, 221
\bibitem{joh78} Johansson, S. 1978  {Phys. Scripta} 18  217

\bibitem{hul82} Huldt, S.; Johansson, S.; Litz\'en U.; Wyart, J.-F 1982 
{Phys. Scripta} 25  401

\bibitem{pic98} Pickering, Juliet C.; Raassen, A. J. J.; Uylings, P. H. M.; Johansson, S. 1998  {Astrophys. J. Sup.} 117  261
\bibitem{ber96} Bergeson, S. D., et al. 1996  {Astrophys. J.} 464  1044

\bibitem{sik99} Sikstr\"om, C. M, et al. 1999  {J. Phys. B: Opt. Atom. Mol. Phys.} 32  568

\bibitem{ros01} Rostohar, D., et al. 2001  {Phys. Rev. Letters}
86  1466

\bibitem{joh77a} Johansson, S. 1977  {Phys. Scripta} 15  183

\bibitem{har03} Hartman, H., et al. 2003  {J. Phys. B: Opt. Atom. Mol. Phys.} 36  197
\bibitem{joh03} Johansson, S., Litz\'en U,; Lundberg, H., Zhang, Z.  2003 {Astrophys. J.} 584 107
\bibitem{joh84} Johansson, S., Jordan, C. 1984  { Mon. Not. Royal Astron. Soc.} 210  239

\bibitem{joh77b} Johansson, S. 1977  { Mon. Not. Royal Astron. Soc.} 178  17

\bibitem{har00} Hartman, H., Johansson, S. 2000  { Astron. \& Astrophys.} 359  627

\bibitem{joh93} Johansson, S., Hamann, F. W. 1993  {Phys. Scripta} T 47  157

\bibitem{bau04} Bautista, M.A. 2004  { Astron. \& Astrophys.} 420  763

\bibitem{bau06} Bautista, M.A., Hartman, H., Gull, T., Smith, N.,Lodders, K. 2006  { Mon. Not. Royal Astron. Soc.} 370  1991

\bibitem{bau07} Bautista, M., et al. 2007  { Mon. Not. Royal Astron. Soc.} 370  1991


\bibitem{ironp} Hummer, D.G., et al. 1993  { Astron. \& Astrophys.} 279  298

\bibitem{bau98} Bautista, M.A. \& Pradhan, A.K.  1998  {Astrophys. J.} 492  650

\bibitem{bau04b} Bautista, M.A., Rudy, R.J. \& Venturini, C.C.  2004  {Astrophys. J.} 604  L129

\bibitem{zha95} Zhang, H.L. \& Pradhan, A.K.  1995 { Astron. \& Astrophys. Sup.} 293  953

\bibitem{ram07} Ramsbotton, C.A., Hudson, C.E., Norrington, P.H., Scott, P.
2007  { Astron. \& Astrophys.} 475  765

\bibitem{zha96} Zhang, H.L.  1996  { Astron. \& Astrophys. Sup.} 119  523

\bibitem{zha97} Zhang, H.L. \& Pradhan, A.K.  1997  { Astron. \& Astrophys. Sup.} 126  373

\bibitem{bau04b} Bautista, M.A. 2004  { Astron. \& Astrophys.} 420  763

\bibitem{bau01} Bautista, M.A. 2001  { Astron. \& Astrophys.} 365  268

\bibitem{mel05} Mel\'endez, M. \& Bautista, M.A.  2005  { Astron. \& Astrophys.} 436  1123

\bibitem{bau97} Bautista, M.A. 1997  { Astron. \& Astrophys. Sup.} 122  167

\bibitem{nah94} Nahar, S.N. \& Pradhan, A.K. 1994  {J. Phys. B: Opt. Atom. Mol. Phys.} 27  429

\bibitem{nah96} Nahar, S.N. 1996  {Phys. Rev. A} 53  1545

\bibitem{bau97b} Bautista, M.A. \& Pradhan, A.K. 1997  { Astron. \& Astrophys. Sup.} 126  365

\bibitem{bau99} Bautista, M.A. 1999  { Astron. \& Astrophys.} 137  529

\bibitem{roo76} Rood, R.T., Steigman, G. \& Tinsley, B.M. 1976  {Astrophys. J.} 207  L57

\bibitem{tos01} Tosi, M. 2001, in "Cosmic Evolution", Eds. E. Vangioni-Flam, R.
Ferlet, \& M. Lemoine (Singapore: Scientific)  77

\bibitem{hog95} Hogan, C.J. 1995  {Astrophys. J.} 441  L17

\bibitem{cha95} Charbonnel, C. 1995  {Astrophys. J.} 453  L41

\bibitem{boo99} Boothroyd, A. \& Sackmann, I.-J. 1999 {Astrophys. J.} 510 232

\bibitem{pal02} Palla, F., Galli, D., Marconi, A., Stanghellini, L., Tosi, M.  2002  {Astrophys. J.} 568 L57

\bibitem{bra02} Brage, T., Judge, P.G. \& Proffit, C.R.  2002 {Phys. Rev. Lett.} 89 281101

\bibitem{rub04} Rubin, R.H., Ferland, G.J., Chollet, E.E. \& Horstmeyer, R. 2004 {Astrophys. J.} 605 784

\bibitem{cas05} Casassus, S., Storey, P.J., Barlow, M.J. \& Roche, P.F.  2005 
{ Mon. Not. Royal Astron. Soc.} 359 1386 

\bibitem{peq94} P\'equignot, D. \& Baluteau, J.-P. 1994 { Astron. \& Astrophys.} 283 593

\bibitem{din01} Dinerstein, H.L. 2001 {Astrophys. J} (Letters) 550 L223

\bibitem{ding01} Dinerstein, H.L. \& Beballe, T.R. 2001  {Astrophys. J.} 562  515

\bibitem{ste05} Sterling, N.C. \& Dinerstein, H.L. 2005, {COSMIC ABUNDANCES as Records of Stellar Evolution and Nucleosynthesis in honor of David L. Lambert}  ASP 336  P. 367

\bibitem{ste07} Sterling, N. C.; Dinerstein, Harriet L.; Kallman, T. R., 2007  {Astrophys. J. Sup.}  169  37

\bibitem{bad04} Badnell, N.R., Berrington, K.S., Summers, H.P., O'Mullane. M.G., Whiteford, A.D. \& Ballance, C.P. 2004  {J. Phys. B: Atom., Mol., Optic. Phys.} 37  4589

\bibitem{ait96} Ait-Tahar, S., Grant, I.P., \& Norrington, P.H.  1996 {Phys. Rev. A} 54 3984

\bibitem{bad03} Badnell, N.R., O'Mullane, M.G., Summers, H.P., Altun, Z., Bautista, M.A. et al. 2003  { Astron. \& Astrophys.} 406  1151 

\bibitem{fog05} Fogle, M., Badnell, N.R., Glans, P., Loch, S.D., Madzunkov, et al.  2005  { Astron. \& Astrophys.} 442   757 

\bibitem{bohm05} B\"ohm, S., M\"uller, A., Schippers, S., Shi, W., Fogle, et al.  2005 { Astron. \& Astrophys.}
37 1151
\bibitem{bad06} Badnell et al.  2006, in "Planetary Nebulae in our Galaxy and Beyond", Eds. M.J. Barlow \& R.H. M\'endez IAUS 234 211

\bibitem{men00} Mendoza, C.  2000, in {Atomic and Molecular Data and their Applications, ICAMDATA. K. A. Berrington and K.L. Bell, eds. American Institute of Physics (AIP) Proceedings, vol. 543. New York} ISBN 1-56396-971-8 p.313

\bibitem{sha03} Sharpee, B., Williams, R., Baldwin, J.A.\& van Hoof, P.A.M.  2003 {Astrophys. J. Sup.} 149 157

\bibitem{bau99b} Bautista, M.A.  1999 {Astrophys. J.} 527 474

\bibitem{bad93} Badnell, N.R., Pindzola, M.S., Dickson,
W.J., Summers, H.P., Griffin, D.C., Lang, J.  1993 {Astrophys. J.} 407 L91

\bibitem{bau02} Bautista, M.A.  2001 { Astron. \& Astrophys.} 365 268

\bibitem{bohm03} B\"ohm, S., M\"uller, A., Schippers, S., Shi, W., Ekl\"ow, N., et al.  2003 { Astron. \& Astrophys.} 405 1157

\bibitem{lee02} Lee, H.-W., Kang, Y.-W. \& Byum, Y.-I.  2001 {Astrophys. J.}
 551 L121

\bibitem{mel06} Bautista, M.A. et al. 2008,
(in preparation)

\bibitem{nah96b} Nahar, S.N. \& Pradhan, A.K.  1996 { Astron. \& Astrophys. Sup.} 119 509

\bibitem{robi} Robicheaux, F. \& Pindzola, 
M.S. 1997 {Phys. Rev. Letters}  79 2237

\bibitem{wil03} Williams, R., Jenkins, E.B., Baldwin, J.A. \& Sharpee, B.  2003 {Pub. Astron. Soc. Pacific} 115 178

\bibitem{bau06b} Bautista, M.A. 2006,  {J. Phys. B: Opt. Atom. Mol. Phys.} 39 L361

\end{thebibliography}
\end{document}